\documentclass{elsart}
\journal{New Astronomy}

\usepackage{natbib}
\usepackage{psfig}
\usepackage{amssymb}

\begin{document}


\newcommand{\n}     {\noindent}%
\newcommand{\sn}    {\smallskip\noindent}%
\newcommand{\ve}    {\vfill\eject}%
\renewcommand{\etal}    {\mbox{et al.\,}}%
\newcommand{\cge}   {\ensuremath{\,\gtrsim\,}}%
\newcommand{\cle}   {\ensuremath{\,\lesssim\,}}%
\newcommand{\bul}   {\ensuremath{\bullet}\ }%
\newcommand{\Msun}  {\ensuremath{M_{\odot}}}%
\newcommand{\zform} {\ensuremath{\mbox{\rm z}_{\mbox{\rm\scriptsize form}}}}%
\newcommand{\arcmpt}    {{$\buildrel{\prime}      \over .$}}
\newcommand{\arcspt}    {{$\buildrel{\prime\prime}\over .$}}
\newcommand{\re}    {\ensuremath{r_e}}%
\newcommand{\micron}    {\ensuremath{\mu}\mbox{\rm m}}%


\begin{frontmatter}

\title {Cosmic infrared background and early stellar populations}

\author{A. Kashlinsky}
\address{SSAI and Code 665, Observational Cosmology Laboratory, Goddard Space Flight Center, Greenbelt, MD 20771\\
 Email:\ kashlinsky@stars.gsfc.nasa.gov}

\begin{abstract}
Cosmic infrared background (CIB) contains information about galaxy
luminosities over the entire history of the Universe and can be a
powerful diagnostic of the early populations otherwise
inaccessible to telescopic studies. Its measurements are very
difficult because of the strong IR foregrounds from the Solar
system and the Galaxy. Nevertheless, substantial recent progress
in measuring the CIB and its structure has been made. The
measurements now allow to set significant constraints on early
galaxy evolution and, perhaps, even detect the elusive Population
III era. We discuss briefly the theory behind the CIB, review the
latest measurements of the CIB and its structure, and discuss
their implications for detecting and/or constraining the first
stars and their epochs.
\end{abstract}


\end{frontmatter}


\section{Introduction}

Cosmic infrared background (CIB) is a repository of stellar
emissions throughout the entire history of the Universe. Cosmic
expansion shifts any photon emitted in the visible/UV bands at
high $z$ into the near-IR (NIR) and the high-$z$ NIR photons will
today appear in mid- to far-IR. Consequently, the NIR part of the
CIB spectrum (1$\mu$m$ < \lambda < $10 $\mu$m) probes the history
of direct stellar emissions from the early Universe, and the
longer wavelengths contain information about the early dust
production and evolution.

The last several years have seen significant progress in CIB
studies, both in identifying/constraining its mean level
(isotropic component) and fluctuations (see \cite{k05} for a
recent review). CIB contains emissions also from objects
inaccessible to current (or even future) telescopic studies and
can, therefore, provide unique information on the history of the
Universe at very early times. One particular example of such
objects, which is of prime relevance to this meeting, are
Population III, the so-far hypothetical zero-metallicity stars
expected to have preceded normal stellar populations seen in the
farthest galaxies to-date.

This review is organized as follows: we briefly present
theoretical basis behind the CIB in Sec. 2, and discuss the two
broad classes of contributors in Sec. 3. In Sec. 4 we review the
old (pre-Spitzer) measurements of the CIB and Sec. 5 gives an
overview of the very recent results that were obtained from
Spitzer IRAC deep exposures by our team \cite{spitzer-pop3}. We
conclude in Sec. 6.

\section{Theory}

The total CIB flux, $F\equiv \lambda I_\lambda$, produced in $dz$
by a population with comoving luminosity function $n(L)$ and
luminosity density ${\mathcal L}(z)=\int n(L)LdL$ is:
\begin{equation}
\frac{dF}{dz} = \frac{cH_0^{-1}}{4\pi} \frac{1}{(1+z)^2}
\frac{d(H_0t)}{dz} \sum_i {\mathcal L}(z) [\lambda
f_\lambda(\frac{\lambda}{1+z}; z)], \label{dfdz}
\end{equation}
Here $f_\lambda$ is the SED of the sources/galaxies normalized to
$\int f_\lambda d\lambda=1$.

Because the emitters are clustered, the CIB produced by them will
have fluctuations reflecting the underlying clustering with power
spectrum $P_3(k)$. The flux fluctuation is $\delta F=F-\langle
F\rangle$ and in the limit of small angles ($<$1 sr) a cartesian
geometry can be used in its Fourier expansion, $\delta
F(\vec{\theta}) = \sum f_q \exp(-i\vec{q}\cdot\vec{\theta})$. The
2-D power spectrum of the CIB is $P_2(q)=\langle |f_q|^2\rangle$
and is related to $P_3$ via the Limber equation. The rms
fluctuations on scale encompassing angular wavelength $2\pi/q$ is
$\delta F_{\rm rms} \simeq \sqrt{q^2P_2/2\pi}$. In order to
connect the measured/projected power spectrum to the underlying
3-D distribution, it is convenient to use the Limber equation in
the form \cite{k05}:
\begin{equation}
\frac{q^2P_2(q)}{2\pi} = \pi t_0 \int \left( \frac{d
I_{\nu^\prime}}{dt}\right)^2 \Delta^2(qd_A^{-1};z) dt
\label{limber}
\end{equation}
where $t_0$ is the time-length of the period over which the CIB is
produced and
\begin{equation}
\Delta^2 (k) = \frac{1}{2\pi^2} \frac{k^2 P_3(k)}{ct_0}
\label{Delta}
\end{equation}
 is the mean-squared fluctuation in number of sources within a line-of-sight cylinder of volume $k^{-2}
 ct_0$. Eqs. \ref{limber},\ref{Delta} can be rewritten in terms of
 the total CIB flux as
\begin{equation}
 \delta F \left(\frac{2\pi}{q}\right) \equiv \sqrt{\frac{q^2P_2(q)}{2\pi}} = F \Delta(q
 d_A^{-1}(\bar{z}))
 \label{effective}
 \end{equation}
 where $\bar{z}$ is a suitably averaged
 effective redshift of the populations contributing to CIB
 fluctuations.

 In addition to fluctuations produced by clustering, there will also be a shot
 noise component due to Poissonian fluctuations from sources occasionally entering the beam.
 This component is small for surveys with large beams, which contain many
 sources, but can be important for $\sim$arcsec or better resolution surveys. The
 power spectrum of the shot noise contribution to the CIB from
 sources with angular density $dN(m)/dm$ per magnitude range $dm$ is:
 \begin{equation}
 P_{\rm SN} = \int f^2(m) \frac{dN}{dm} dm
 \label{shotnoise}
 \end{equation}
 where $f(m)\equiv f_010^{-0.4m}$ is the flux corresponding to
 magnitude $m$.

 Any measurement of the CIB power spectrum is affected by the
 sample or cosmic variance. This is a consequence of the fact that
 one is trying to deduce the power spectrum of an infinite
 distribution from a field of finite extent. The uncertainty is
 $N_q^{-1/2}$ where $N_q$ is the number of independent elements in Fourier
 space that goes into determining the power at given $q$. Because
 at large scales (small $q$) there are progressively fewer such
 elements, reaching just one at half the field extent, in order to
 reliably probe fluctuations on scale $\theta$ one needs a field-of-view a
 few times larger.

 In surveys with high angular resolution it becomes possible to
 remove bright(er) more nearby galaxies. In this way one can
 isolate the contributions to the CIB from fainter galaxies
 located at progressively higher $z$ \cite{komsc,okmsc}.

\section{CIB contributors}

Galactic and Solar System foregrounds are the major obstacles to
space-based measurements of the CIB. Galactic stars are the main
contributors at near-IR ($<$ a few micron), zodiacal light from
the dust in the Solar system contributes mostly between $\sim$10
and $\sim$50 micron, and Galactic cirrus emission produces most of
the foreground at IR wavelengths longward of $\sim 50$ micron.
Accurately removing the foregrounds presents a challenge and many
schemes have been developed to do this as well as possible. Stars
can be removed in surveys with fine angular resolution
\cite{gorjian,komsc,okmsc} or by statistically extrapolating the
various stellar contributions to zero \cite{paper3,wright_reese}.
Zodiacal light contributions usually are removed using the DIRBE
zodi model \cite{kelsall} or its derivatives \cite{gorjian,irts}.
Galactic cirrus and zodiacal light are both intrinsically diffuse,
but are fairly homogeneous adding to the usefulness of CIB
fluctuations studies at mid- to far-IR \cite{paper2}.

For purposes of this review we divide the extragalactic CIB
contributors into two groups: 1) ``ordinary" galaxies, which are
metal-enriched systems of normal stellar populations with Salpeter
of Scalo IMFs, and 2) Population III stars that preceded ordinary
galaxy populations. The first category, which we term ordinary
galaxies contribute around $\sim$1-10 nW/m$^2$/sr to the mean CIB
in the near- to mid-IR. At many wavelengths their total flux can
be measured by summing up contributions from deep galaxy counts,
$\int f(m) \frac{dN}{dm} dm$, and several surveys go sufficiently
deep to observe this quantity saturate. In this way one can arrive
at a fairly robust estimate of the total contribution of the
ordinary galaxies to the CIB. Such contributions in near-IR are
shown in Fig. 1 with open squares for HDF data
\cite{madaupozzetti} and with open diamonds for Spitzer/IRAC deep
counts \cite{fazio}.

Population III are currently believed to have been very massive
stars, as discussed by Tom Abel in these proceedings, but even
then their contributions can only be estimated in a theoretical,
and necessarily speculative, manner. However, if the CIB
measurements are accurate and the net contribution from ordinary
galaxies is known to a reasonable accuracy, the difference between
the two is likely to originate from Population III stars located
at very early cosmic times ($z>10$ or so). Even the most careful
measurements of the CIB mean levels can be significantly affected
by the systematics and be mistaken for the various residual
errors. On the other hand, Population III stars, whose emission
arises at epochs when the spatial spectrum of galaxy clustering is
not yet evolved, should have produced a unique and measurable
signature in the near-IR CIB fluctuations \cite{cooray,kagmm}.
That signature, both its spectrum in the angular and energy
frequency domain, could provide - if and when measured - the
ultimate insight into the Population III epochs. There are several
intuitive reasons why Population III produce significant CIB
levels, both its mean and fluctuations:\\ $\bullet$ Each unit mass
of Population III (if made of massive stars) emits $\sim 10^5$
more light than normal stars and massive stars (assuming
Population III were indeed massive stars) radiate with a higher
mean radiative efficiency, $\epsilon =0.007$, than the present day
stellar populations, leading to substantial net flux.\\ $\bullet$
Because their era was presumably brief, Population III epochs
contain less projected volume than the ordinary galaxy populations
spanning the epochs of Population I and I stars. Hence, the larger
relative fluctuations.\\ $\bullet$ Biasing is higher for
Population III because they form out of rarer regions which leads
to the amplified correlations.

\section{``Old" results: DIRBE, FIRAS, IRTS, 2MASS}

The current best direct measurements of the mean level of the CIB
come from the data obtained with the COBE DIRBE and FIRAS
instruments and the IRTS mission. The current measurements of the
CIB anisotropies come from the above missions, as well as the
ground based deep 2MASS images and the very latest Spizter
results. The latter are discussed separately in the next section.

\n\makebox[\textwidth]{
   \psfig{file=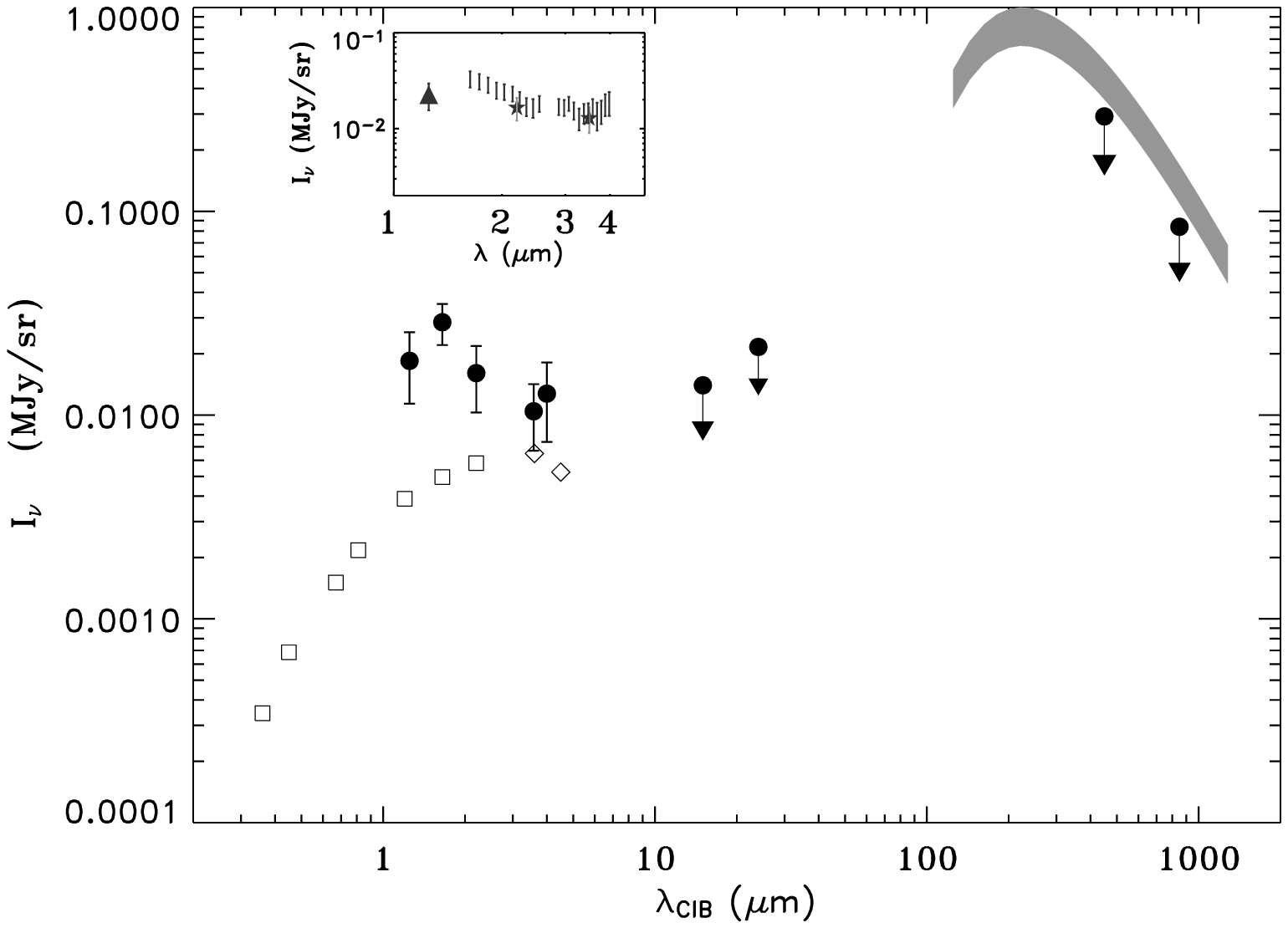,width=\textwidth,angle=0}\ \
}

\n {\footnotesize\baselineskip=10pt {\bf Figure~1:}\ Inset shows
the near-IR CIB detections with their error bars: filled triangle
is from \cite{cambresy} based on DIRBE data, vertical error bars
from IRTS data \cite{irts}, and asterisks  are from analyses based
on DIRBE data \cite{dwekarendt,gorjian,wright_reese}. Shaded
region corresponds to the far-IR CIB based on FIRAS data
\cite{puget} with a fit and uncertinties from \cite{firas}. Open
symbols at visible to near-IR correspond to net contributions from
ordinary galaxies: those derived from the HST data are shown with
squares \cite{madaupozzetti} and those from Spitzer data are shown
with diamonds \cite{fazio}. The difference is the CIB excess over
contributions from ordinary galaxies which is shown with filled
circles marked with downward pointing arrows where galaxies can
account, within errors, for the CIB and only upper limits
(essentially errors) can be claimed \cite{k05}. }

In the near-IR detections are difficult because of the substantial
foreground by Galactic stars. Claims of detections of the mean
isotropic part of the CIB are based on various analyses of DIRBE
and IRTS data
\cite{dwekarendt,irts,gorjian,wright_reese,cambresy,arendtdwek}.
The measurements agree with each other, although the methods of
analysis and foreground removal differ substantially. They also
agree with the measured amplitude of CIB fluctuations discussed
later in this section. The results seem to indicate fluxes
significantly in excess of those from observed galaxy populations.
The near-IR detections of the CIB are summarized in the inset of
Fig. 1. Total fluxes from galaxies in NIR from HST and Spitzer
IRAC measurements are shown with open symbols in Fig. 1; they
saturate at Vega magnitude of about K$\simeq 20-22$ and are a
factor is $\sim 2-3$ below the detected CIB. The difference
between the claimed CIB levels and the total fluxes from
``ordinary" (i.e. not Population III) galaxies are shown with
their uncertainties as filled circles in Fig. 1.

In mid-IR there are no CIB detections, but the best {\it upper}
limits come from gamma-ray absorption limits derived from spectra
of nearby ($z\sim 0.1$) blazars and are quite low $\sim 4-5$
nw/m$^2$/sr \cite{stanev}. Because the foregrounds are very
homogeneous at these wavelengths, one obtains slightly higher, but
comparable, CIB upper limits ($<\atop\sim $10 nW/m$^2$/sr) from
the fluctuations analysis \cite{paper2}. The best upper limits on
the CIB at mid-IR are comparable to the net contributions from ISO
\cite{metcalfe} and Spitzer MIPS \cite{papovich} galaxy counts of
$\simeq 3$ nW/m$^2$/sr at 15 and 24 $\mu$m.

Far-IR CIB measurements come from the COBE FIRAS
\cite{puget,firas} and COBE DIRBE \cite{schlegel,hauser} data
analysis. They are broadly consistent although the COBE DIRBE
detections at 140 and 240 $\mu$m give larger fluxes. The fit to
the FIRAS CIB detection of \cite{firas} is shown within its
uncertainties as the shaded region in Fig. 1. The far-IR CIB is
expected to originate from dust (re)emissions and the shape of the
FIRAS detection in Fig. 1 confirms this. The total galaxy fluxes
there come from SCUBA galaxies \cite{blain} and, within the
uncertainties, they can account for the CIB at these wavelengths.

Detections of CIB fluctuations come from three independent
experiments and are consistent with each other if accounting is
made for contributions from removed galaxies, such as in deep
2MASS data \cite{okmsc}. The DIRBE data based analysis,
extrapolating the contribution from Galactic stars to 0, gives a
statistically significant CIB fluctuation at $\sim0.5^\circ$ in
the DIRBE first 4 channels or 1.25, 2.2, 3.5 and 4.5 $\mu$m
\cite{paper3}. The IRTS team \cite{irts} measured the spectrum of
CIB anisotropies on degree scales from IRTS data band-averaged to
$\sim $ 2 $\mu$m; their results are consistent with \cite{paper3}.
Because the beam was large in both instruments, no galaxy removal
was possible in the data and the CIB anisotropies arise from {\it
all} galaxies, i.e. from $z=0$ to the earliest times. Based on the
amplitude of the present-day galaxy clustering, the NIR detections
exceed the expectations from galaxies evolving with no or little
evolution by a factor of $\sim 2-3$ and are consistent with the
measured (high) mean CIB levels in the NIR.

\n\makebox[\textwidth]{
   \psfig{file=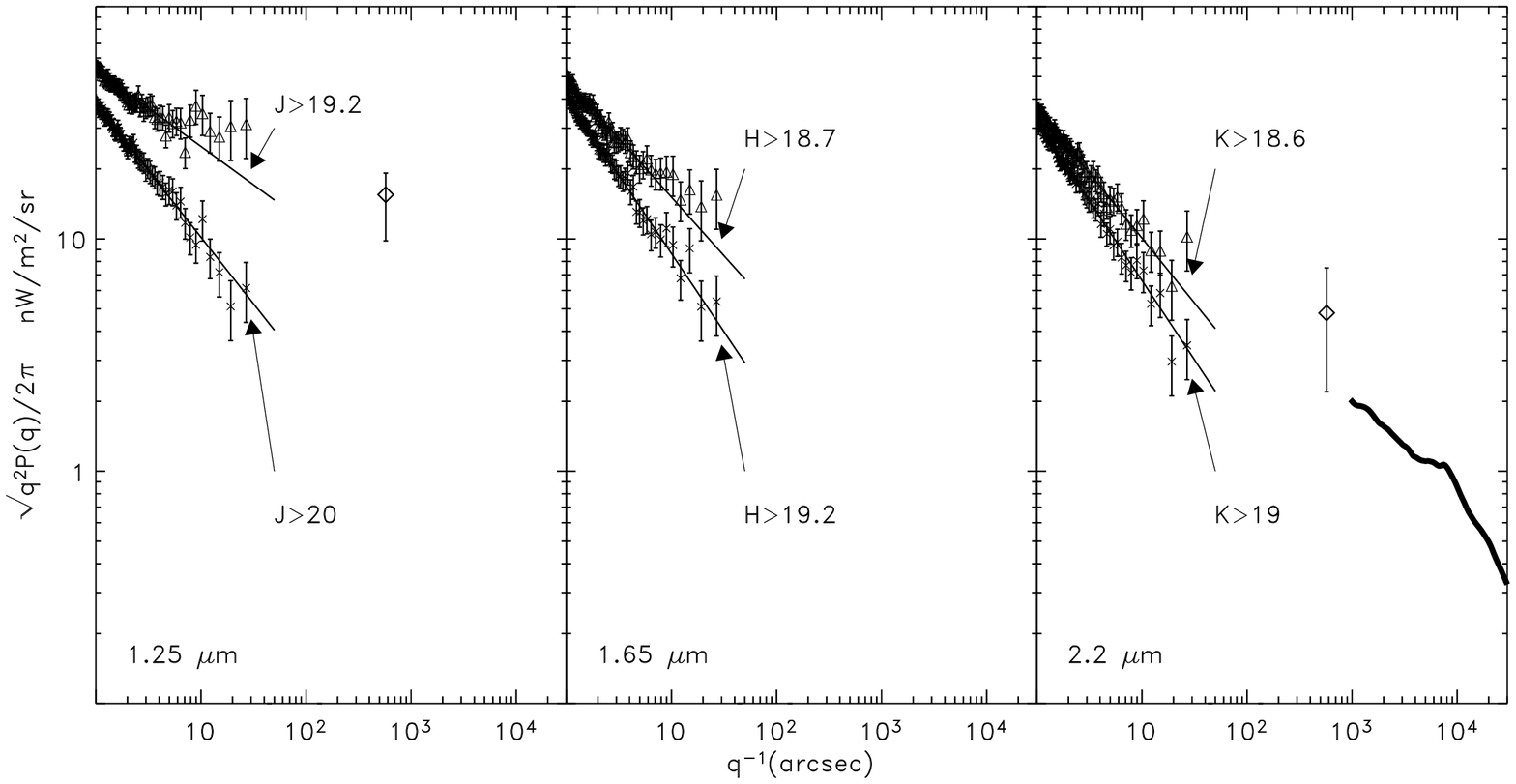,width=\textwidth,angle=0}\ \
}

\n {\footnotesize\baselineskip=10pt {\bf Figure~2:} CIB
fluctuations at J, H and K bands: open triangles and crosses show
CIB fluctuations arising from galaxies fainter than the limit
marked near the least squares fit lines based on deep exposure
2MASS data \cite{komsc,okmsc}. In the right panel for 2.2 $\mu$m
the open diamond with the 92\% confidence level error shows
detection based on DIRBE data from \cite{paper3}. The thick solid
line shows the CIB fluctuation spectrum based on the IRTS data
from \cite{irts}. Both of the latter correspond to CIB from all
galaxies; the low resolution of the DIRBE and IRTS/NIRS
instruments did not allow for galaxy removal. \ }

Analysis of deep 2MASS data \cite{komsc,okmsc}  took the
measurements of the CIB anisotropies one step further in that it
has allowed to remove galaxies out to $K\simeq (18.5-19)$ and
measure the spectrum of CIB anisotropies from {\it fainter}
galaxies in the J, H, and K 2MASS bands on sub-arcminute angular
scales. These galaxies are typically located at cosmological times
when the Universe was less than $\sim$ half its present age so
this measurement isolates CIB ansitropies arising from the
emissions in the Universe at these early times. When extrapolated
to the present-day the 2MASS based results are consistent with the
DIRBE- and IRTS-based measurements. Fig. 2 summarizes the
2MASS-based results and compares them with the other measurements.

If one were to remove galaxies to fainter limits than was possible
in the 2MASS data, one can probe emissions arising at still
earlier cosmic times. There is a possibility - of direct relevance
to the topics of this meeting - of Population III being
responsible for the CIB excess \cite{santos,salvaterra}. In order
to directly measure such a contribution it is necessary to remove
ordinary galaxies down to very faint flux limits. This was
impossible to do with the old instruments, which either had low
resolution and sensitivity (DIRBE and IRTS/NIRS) or were
ground-based (2MASS) and affected by the atmosphere. We discuss a
recent Spitzer-based measurement of the CIB from early epochs in
the next section.

\section{New results from Spitzer IRAC deep images}

There is no direct observational evidence on what the first stars,
Population III, were, although it is (now) generally believed that
they had to be very massive stars. Recently, two teams have
suggested that Population III, if massive, should have left
substantial CIB fluctuations with a spectrum serving as their
signature \cite{cooray,kagmm}. Uncovering these anisotropies is a
difficult and challenging task, but it offers a concrete
observational test of and probe into the Population III era. We
(Kashlinsky, Arendt, Mather, Moseley) have attempted to make a
measurement of these CIB fluctuations in deep images obtained with
Spitzer IRAC at 3.6, 4.5, 5.8 and 8$\mu$m. This analysis is
presented in a manuscript forthcoming in Nature
\cite{spitzer-pop3} to which we refer readers for details; below
is a brief summary of the steps in the measurement and the main
results.

The main data came from the IRAC GTO observations with $\sim 8-9$
hour integration of a field located at sufficiently high Galactic
latitude ($b_{\rm Gal} =36.1^\circ$). Additionally, we analyzed
shallower observations for 2 auxiliary fields in order to test for
isotropy of any cosmological signal. The datasets were assembled
out of the individual frames using the least calibration method of
\cite{fixsen}. This self-calibration procedure has advantages over
the standard pipeline calibration of the data in that the derived
detector gain and offsets match the detector at the time of the
observation, rather than at the time of the calibration
observations. The most noticeable difference is that the different
AORs are affected by different patterns of residual images left by
prior observations, which if not properly removed, can leave
artifacts in the final images \cite{longpaper}.

For the final analysis we selected a subfield of $\simeq 5^\prime
\times 10^\prime$ with a fairly homogeneous coverage. After the
image assembly we removed a liner gradient from each map.
Individual sources have been clipped iteratively as positive flux
excursions above a fixed threshold. We used the full-array PSF in
removing the residual extended parts of the identified sources by
a CLEAN-type procedure. Fig. 3 shows the final image for Channel 1
(3.6 $\mu$m) of IRAC. This image has $> 75\%$ pixels left which
allows robust computation of the diffuse flux Fourier transforms,
$f_q$. CIB fluctuations from Population III located at very high
$z$ should be independent of the clipping threshold, so we also
clipped progressively deeper to verify that our results are
threshold-independent. As more pixels are removed, it becomes
impossible to evaluate robust Fourier transforms, so we calculated
instead the diffuse flux correlation function, which is related to
the power spectrum by an integral transform. The power spectrum
was evaluated as $P(q) = \langle |f(q)|^2\rangle$ where the
average is taken over narrow concentric rings around a central
$q$. Instrument noise was evaluated by dividing the data into two
subsets (A,B) containing the odd and even numbered frames and
evaluating the power spectrum of the $\frac{1}{2}(A-B)$ map.

\n\makebox[\textwidth]{
   \psfig{file=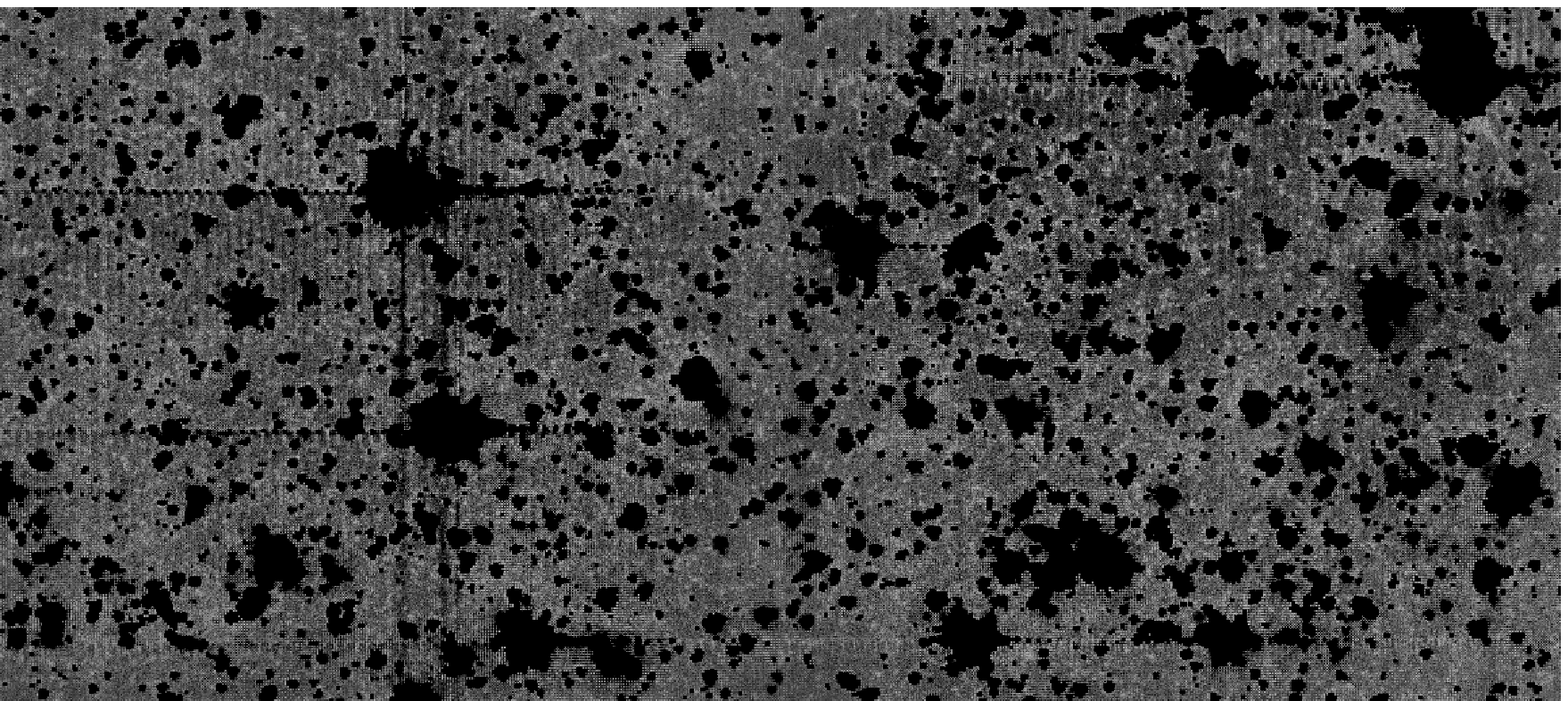,width=\textwidth,angle=0}\ \
}

\n {\footnotesize\baselineskip=10pt {\bf Figure 3}\ Final map for
analysis from \cite{spitzer-pop3} at 3.6 $\mu$m. The map has point
sources subtracted and clipped. This positive image is scaled from
[-2.0,4.0] nW m$^{-2}$ sr$^{-1}$. The regions of clipped sources
are indicated by black area which are set to 0.0 for further
analysis.\ }

In this analysis we detected fluctuations significantly exceeding
the instrument noise. Fig. 4 shows the CIB fluctuation (the
instrument noise has been subtracted) for the IRAC Channel 1 at
3.6 $\mu$m. The excess fluctuation on arcminute scales in the 3.6
$\mu$m channel is $\sim 0.1$ nW/m$^2$/sr; our analysis shows a
similar amplitude in the longer IRAC bands indicating that the
energy spectrum of the arcminute scale fluctuations is flat to
slowly rising with increasing wavelength at least over the IRAC
range of wavelengths.

The signal is significantly higher than the instrument noise and
the various systematics effects we investigated cannot account for
it. We find a statistically significant correlation between the
channels for the region of overlap. We also find that the
correlation function evaluated at deeper clipping cuts, when we
have too few pixels left for Fourier analysis, remains the same
and is consistent with the power spectrum evaluations. This
suggests that the signal comes from the sky, which contains
contributions from the remaining suspects: zodiacal emission,
Galactic cirrus, and the extragalactic components: the ``ordinary"
galaxies and the putative Population III.

The zodiacal and Galactic cirrus components are shown in Fig. 4
and are much smaller than the signal and have different energy
spectrum than these foregrounds, although we find that Channel 4
(8 $\mu$m) data may contain a significant cirrus contribution. The
latter cannot account for all the signal since the maps common to
all four channels have statistically significant correlation and
the cirrus emission is negligible at shorter wavelengths.

Contributions from ordinary galaxies come in two flavours: 1) the
shot noise, and 2) due to clustering out of the primordial density
fields which these galaxies trace. The shot-noise contribution,
when evaluated directly from galaxy counts, gives a good fit to
the fluctuations on smaller scales. Ordinary galaxies have been
eliminated from the maps down to very faint flux levels ($\sim
0.3\mu$Jy in Channel 1). The remaining galaxies' contribution to
the CIB mean levels is small ($<0.1-0.2$ nW/m$^2$/sr) and in
\cite{spitzer-pop3} we showed that their clustering component is
much smaller than the signal in Fig. 4. On the other hand, the
amplitude, power spectrum and the spectral energy dependence of
the $>$ arcminute-scale fluctuation can be explained by emissions
from Population III \cite{spitzer-pop3}.

\n\makebox[\textwidth]{
   \psfig{file=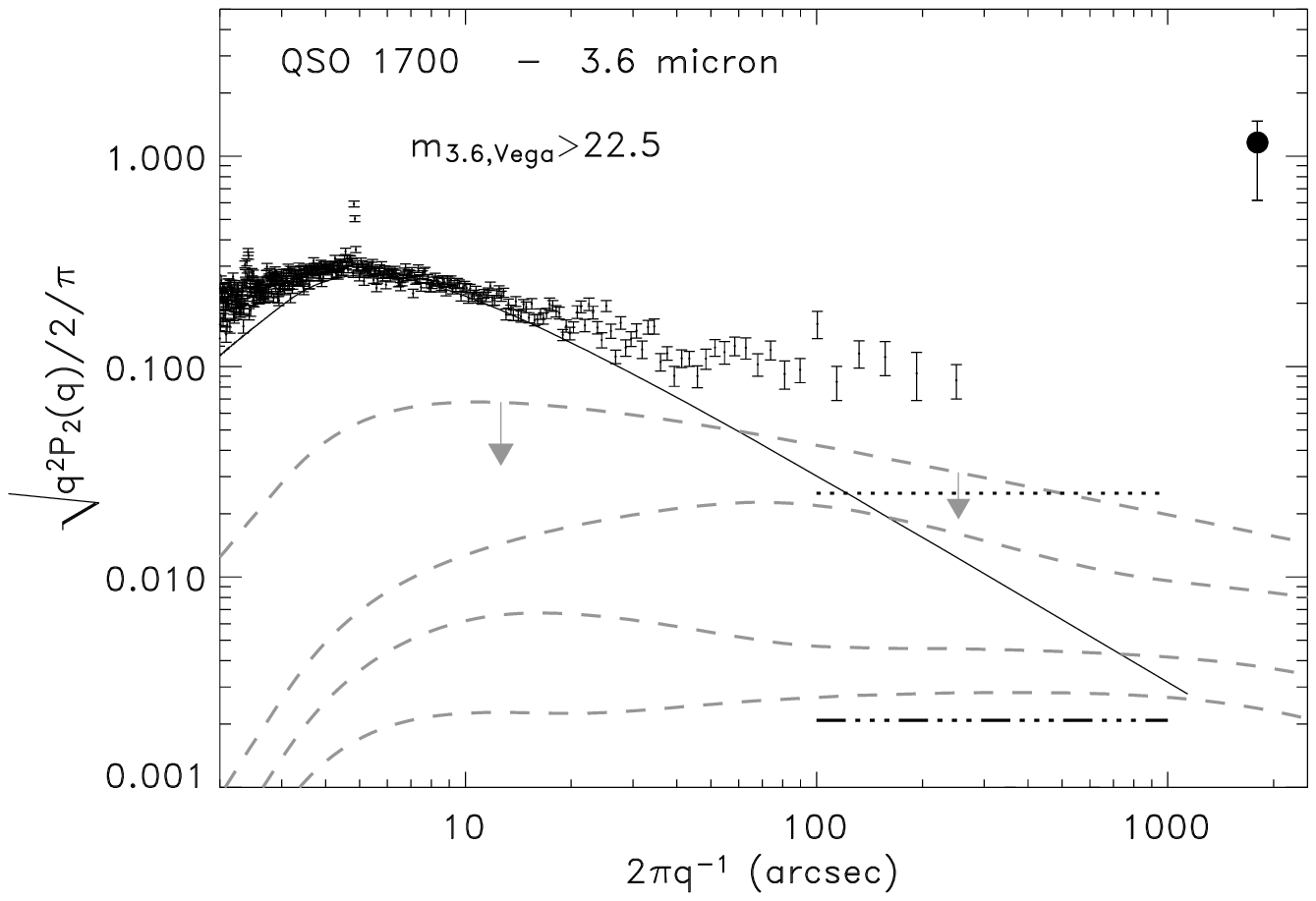,width=\textwidth,angle=0}\ \
}

\n {\footnotesize\baselineskip=10pt {{\bf Figure 4} The spectrum
of CIB fluctuations in the IRAC Channel 1 is shown with error
bars. Solid line shows the shot noise contribution from ordinary
galaxies. Contributions from the clustering component are shown
with dashed lines, but are small: the top dashed line shows the
upper limit assuming that galaxies at early times remained
clustered as today, and the other dashed lines correspond to the
remaining galaxies with flux less 0.3 $\mu$Jy lying at $z\geq $1,
3 and 5 respectively. Cirrus and zodiacal emission contributions
to the fluctuations are shown with dotted and dashed-dot-dot-dot
lines respectively. The result at a larger scale from the DIRBE
analysis \cite{paper3}, which includes contributions from all
galaxies, is shown as filled circle with its 92\% confidence
level. }\ \label{fig_ch1}}

The results are easier to interpret in terms of the high-$z$
Universe than the earlier DIRBE, deep 2MASS and NIRS data analyses
discussed in Sec. 4. The latter claimed significant CIB levels in
the NIR, both the mean levels
\cite{dwekarendt,irts,wright_reese,gorjian,cambresy} and
anisotropies \cite{paper3,irts,komsc,okmsc}, which are higher than
the contributions from the observed ordinary galaxy populations,
which saturate with the fainter galaxies contributing little to
the total flux. However, any direct comparison between the old
fluctuation measurements and the present detections is difficult
because of the different range of angular scales, wavelengths, and
because in the DIRBE \cite{paper3} and NIRS analyses \cite{irts}
no foreground galaxies were removed, and the deep 2MASS data also
contained contributions from relatively bright galaxies. What is
important is that the new measurement identifies directly the CIB
fluctuations from very faint and, hence, distant sources and on
scales $>1^\prime$ the fluctuations seem to be dominated by
Population III emissions.

\section{Conclusions}

Cosmic infrared background often presents the only way to detect
emissions from all cosmic times including from objects
inaccessible to current or future telescopic studies. Because of
the strong foreground emissions from the Galaxy and the Solar
system, even in space-based measurements one often has to get at
the CIB indirectly. On the other hand, in surveys with fine
resolutions and good sensitivity one can remove galaxies down to
sufficiently faint fluxes to start isolating contributions to the
CIB and its structure from objects at progressively earlier times.
If enough ordinary galaxies are removed, one can hope to see the
contribution to the CIB from Population III era. We discussed the
latest measurements of the CIB and its fluctuations and the
constraints it set on early stellar populations. Our recent
analysis of deep data from the Spitzer Space Telescope has allowed
to remove galaxies to very faint levels of $\sim 0.3\mu$Jy. We
find in the remaining data CIB anisotropies whose amplitude
exceeds significantly that produced by the remaining galaxies and
probably comes from the Population III epochs \cite{spitzer-pop3}.

\section*{Acknowledgements} This work was supported by the NSF under
Grant No. AST-0406587. I thank my collaborators on the Spitzer
results presented in Sec. 5, Rick Arendt, John Mather and Harvey
Moseley, for their contributions.

\n


\begin{thebibliography}{99}
\frenchspacing
{\footnotesize
\baselineskip=8pt
\bibitem {arendtdwek} Arendt, R.G. \&
Dwek, E. 2003, Ap.J., 585,305\\
\bibitem {longpaper} Arendt, R.G., Kashlinsky, A., Mather, J.C. \&
Moseley, S.H. 2005, Ap.J., in preparation\\
\bibitem {blain} Blain, A. et al 2002, Physics Reports, 369, 111\\
\bibitem {cooray} Cooray, A., Bock, J., Keating, B., Lange, A. \& Matsumoto, T.
2004, 606, 611\\
\bibitem {cambresy} Cambresy, L., Reach, W. T., Beichman, C.A.
\& Jarrett, T. H. 2001, Ap.J., 555, 563\\
\bibitem {dwekarendt} Dwek, E. \& Arendt, R. 1998,
Ap.J.,508,L9\\
\bibitem {fazio} Fazio, G. G. et al 2004, Ap.J.Suppl., 154, 10\\
\bibitem {firas} Fixsen, D. et al. 1998, Ap.J.,508, 123\\
\bibitem {fixsen} Fixsen, D., Moseley, S.H. \& Arendt, R.G. 2000,
Ap.J. Suppl., 128, 651\\
\bibitem {gorjian} Gorjian, V., Wright, E.L., \& Chary, R.R.
2001,Ap.J.,536,550\\
\bibitem {hauser} Hauser, M. et al. 1998, Ap.J., 508, 25\\
\bibitem{k05} Kashlinsky, A. 2005, Physics Reports, 409, 361 \\
\bibitem{spitzer-pop3} Kashlinsky, A., Arendt, R.G., Mather, J.C.
\& Moseley,  S.H. 2005, Nature, in press\\
\bibitem {paper2} Kashlinsky, A., Mather, J. \& Odenwald, S. 1996,
Ap.J., 479, L9\\
\bibitem {paper3} Kashlinsky, A. \& Odenwald, S. 2000a,
Ap.J.,528,74\\
\bibitem {komsc} Kashlinsky, A., Odenwald, S., Mather, J.C.,
Skrutskie, M. \& Cutri, R. 2002, Ap.J,579,L53\\
\bibitem {kagmm} Kashlinsky, A., Arendt,
R., Gardner, J.P., Mather, J. \& Moseley, S.H. 2004, Ap.J., 608,
1\\
\bibitem {kelsall} Kelsall, T. et al 1998, Ap.J., 508,44\\
\bibitem {madaupozzetti} Madau, P. \& Pozzetti, L.
2000,MNRAS,312,L9\\
\bibitem {irts} Matsumoto, T. et al
2005, Ap.J., 626, 31\\
\bibitem {metcalfe} Metcalfe, L. et al. 2003, Astron. Astrophys.,
407, 791\\
\bibitem {okmsc} Odenwald, S., Kashlinsky, A., Mather, J.C.,
Skrutskie, M. \& Cutri, R. 2003, Ap.J,583,535\\
\bibitem {papovich} Papovich, C. et al. 2004, Ap.J. (Suppl.), 154,
70\\
\bibitem {puget} Puget, J.-L. et al. 1996, Astron. Astrophys.,
308, L5\\
\bibitem {salvaterra} Salvaterra, R. \& Ferrara, A. 2003, MNRAS,
339, 973\\
\bibitem {santos} Santos, M., Bromm, V. \& Kamionkowski, M. 2002,
MNRAS, 336, 1082\\
\bibitem {schlegel} Schlegel, D., Finkbeiner, D. P. \& Davis, M.
1998, Ap.J., 500, 525\\
\bibitem {stanev} Stanev, T. \& Franceschini, A. 1998, Ap.J., 494,
L159\\
\bibitem {wright_reese} Wright, E.L. \& Reese, E.D. 2000,
Ap.J.,545,43\\
 }
\end{thebibliography}
\end{document}